\documentclass[journal=jacsat,manuscript=article]{achemso}
\usepackage{chemformula} % Formula subscripts using \ch{}
\usepackage[T1]{fontenc} % Use modern font encodings
\usepackage{graphicx}% Include figure files
\usepackage{dcolumn}% Align table columns on decimal point
\usepackage{bm}% bold math
\usepackage{comment} %comment out blocks of code

\makeatletter
\newcommand*{\forcekeywords}{
  \acs@keywords@print
  \let\acs@keywords@print\relax
}
\makeatother

\author{Bob Minyu Wang}
\author{Yuqing Zhu}
\author{Henry Clark Travaglini}
\author{Sergey Y. Savrasov}
\author{Dong Yu}
\email{yu@physics.ucdavis.edu}
\affiliation[University of California, Davis]
{Department of Physics and Astronomy, University of California, Davis}

\title[]{Schottky Electric Field Induced Circular Photogalvanic Effect in Cd\textsubscript{3}As\textsubscript{2} Nanobelts}


\begin{document}

\begin{tocentry}
\includegraphics[scale=0.47]{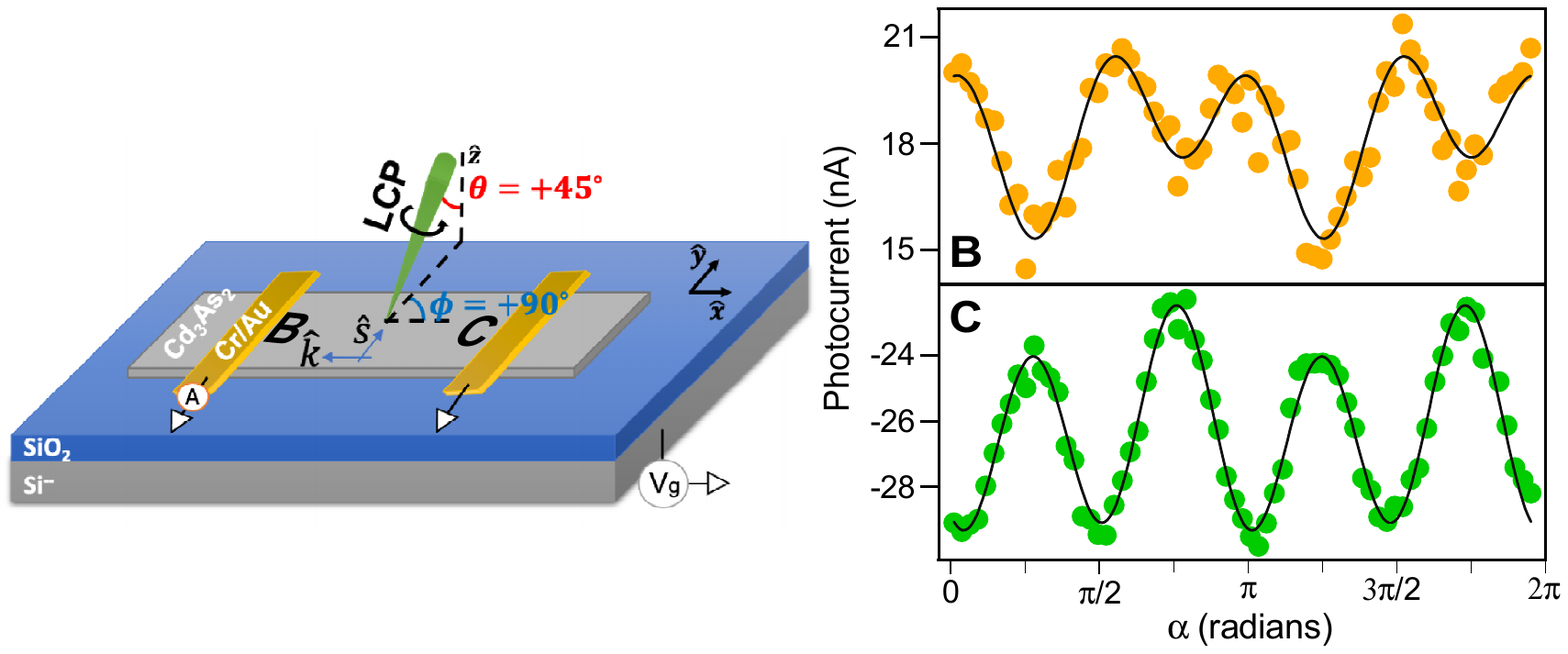}

\end{tocentry}

\begin{abstract}

Dirac semimetals are expected to forbid the manifestation of the circular photogalvanic effect (CPGE) because of their crystal inversion symmetry. Here, we report the observation of the CPGE in Cd\textsubscript{3}As\textsubscript{2} nanobelt field effect transistors, when the photoexcitation is focused in the vicinity of the metal contacts up to room temperature. We attribute the CPGE to the Schottky electric field induced symmetry breaking, which results in the photocurrent modulation by circularly polarized photoexcitation via spin-momentum locking. The hypothesis is supported by a suite of experiments including spatially and angularly resolved helicity dependent photocurrent, Kelvin probe force microscopy, and gate voltage dependence. First principles calculations confirmed a topological phase transition upon field induced structural distortion. This work provides key insights on the electrically controlled helicity dependent optoelectronics in Dirac materials.

\end{abstract}

\forcekeywords

\newpage

The circular photogalvanic effect (CPGE) has gained traction recently as a means to study spin dependent carrier transport in crystals possessing spin-split bands \cite{liu2020circular} and spin-momentum locked topological surface states (TSS)~\cite{mciver2012control, pan2017helicity}. This requires breaking of crystal inversion symmetry or time reversal symmetry. For example, the inherent noncentrosymmetry of the lattice can satisfy inversion symmetry breaking, such as in the Weyl semimetal (WSM) TaAs~\cite{yang2015weyl}. Crystal symmetry can also be lowered at a terminating surface as in the cases of the Rashba spin-split bands of GaAs \cite{ganichev2000circular} and of inorganic-organic lead halide perovskites,~\cite{wang2020spin} and the TSS of topological insulators (TIs) such as Bi$_2$Se$_3$~\cite{mciver2012control}. Furthermore, in centrosymmetric materials, applied electric fields \cite{du2021engineering,dhara2015voltage} and lattice strain \cite{liang2022strain} can break crystal inversion symmetry, and magnetic fields \cite{piatrusha2019topological, baidya2020first} or ultrafast pumping of circularly polarized light can break time reversal symmetry\cite{du2021engineering,wang2013observation}. CPGE not only offers a valuable tool for examining spin-orbit interaction in materials, but also provides an optical control of charge transport applicable to spintronic and quantum devices. Recently, it has also been theoretically proposed that the CPGE induced current may be quantized to a material-independent value, similar to quantized conductance, potentially providing a direct detection of the topological charge of Weyl points~\cite{de2017quantized}. 

Fundamentally, the CPGE relies on asymmetric contributions to photocurrent from incident helical photons and is dictated by angular momentum selection rules. In the case of the Dirac materials that we focus on in this letter, excitations typically involve the Dirac cones with linear dispersion in the vicinity of the Dirac point that exhibit strong coupling between spin and momentum~\cite{hsieh2009tunable}. Net helicity dependent photocurrent is induced if there exists an odd number of Dirac cones in the Brillouin zone as in TIs \cite{mciver2012control} or if contributions from Weyl nodes do not cancel as in WSMs~\cite{yang2015weyl}. In contrast, if a crystal obeys both time reversal and crystal inversion symmetry, as in a Dirac semimetal (DSM), the CPGE is not expected to manifest as Weyl nodes of opposite chirality are paired up at the same position in k-space and any helical contributions to photocurrent will cancel. However, the CPGE can be achieved in these materials via the addition of perturbations such as strain or electric field to systematically break the inversion symmetry~\cite{wang2013three}. The electric control is particularly attractive, as it offers unique opportunities for quantum applications of Dirac materials with in-situ tunability. Electric field induced topological phase transitions have been demonstrated in ultrathin Na\textsubscript{3}Bi films~\cite{collins2018electric}. Schottky electric field induced CPGE has been demonstrated previously in semiconductor nanowires, WSMs, and 2D transition metal dichalcogenides. Their proposed mechanisms include the electron orbital mixing in Si nanowires~\cite{dhara2015voltage}, the Fermi level tilting in TaIrTe\textsubscript{4}~\cite{ma2019nonlinear}, and the symmetry reduction and Berry phase in monolayer MoS\textsubscript{2}~\cite{quereda2021role}. 

Cd\textsubscript{3}As\textsubscript{2} is a prototypical DSM with chemical stability and unusually high carrier mobility~\cite{jay1977electron}. This material system exhibits a plethora of exotic fundamental phenomena such as 3D quantum Hall effect~\cite{zhang2019quantum}, Fermi arc spin transport~\cite{lin2020electric}, long spin coherence length~\cite{stephen2021room}, and giant magnetoresistance~\cite{liang2015ultrahigh}. In addition to basic science, promising applications such as fast broadband photodetectors \cite{wang2017ultrafast} and topological electronics \cite{tokura2017emergent} have been demonstrated using Cd\textsubscript{3}As\textsubscript{2} and related semimetals. Understanding spin-dependent charge transport in this system is paramount to utilizing its capabilities. Though lattice strain-induced CPGE in Cd\textsubscript{3}As\textsubscript{2} thin films has been experimentally demonstrated~\cite{liang2022strain}, manipulation of TSS and bulk band structure via electric fields in DSMs has only been proposed theoretically in Cd\textsubscript{3}As\textsubscript{2}~\cite{baba2019electric, pan2015electric}. Here, we experimentally demonstrate helicity dependent photocurrent (HDPC) observed at the interface made by Cd\textsubscript{3}As\textsubscript{2} and the metal contact, where crystal inversion symmetry is broken via the Schottky electric field, confirmed by surface potential measurements. The degree of helicity dependence can be controlled by the electric field strength through a gate voltage. First principles calculations further support that the DSM can be tuned into a TI under the field induced distortion to the crystal structure. 

\begin{figure}
    \centering
    \includegraphics[width=15cm]{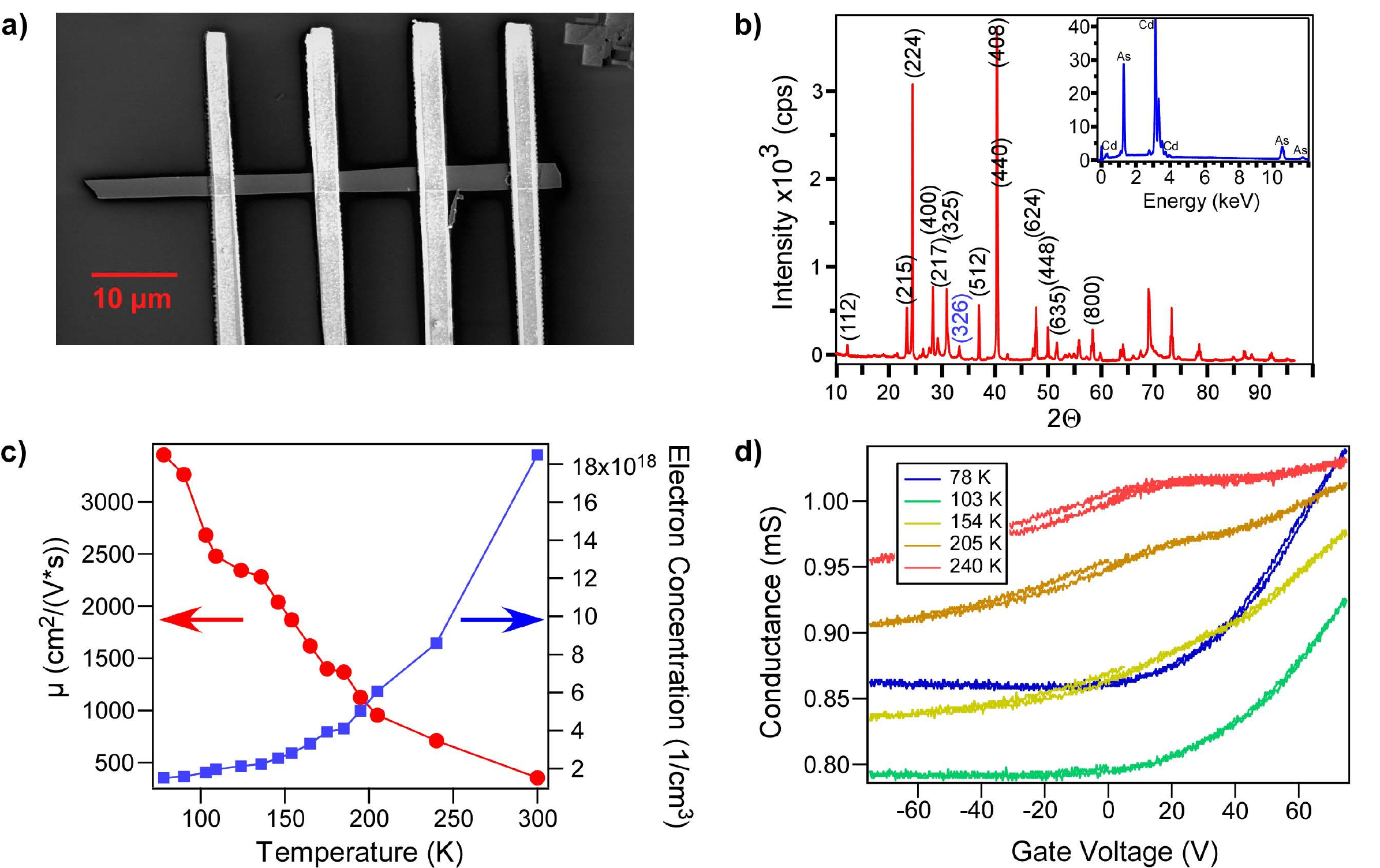}
    \caption{Cd\textsubscript{3}As\textsubscript{2} nanobelts and device characteristics. (a) Scanning electron microscopic (SEM) image of a representative nanobelt FET. (b) Powder XRD spectrum of Cd\textsubscript{3}As\textsubscript{2} nanobelts on an as-grown substrate, in agreement with the (142) phase. A small peak indicated by (326) in blue implies the existence of a mixed phase (133). Inset: EDS of the sample. (c) Electron mobility and concentration versus temperature, extracted from field effect measurements. (d) Conductance versus gate voltage at various temperatures. (c) and (d) were taken in Device \#1.}
    \label{Figure 1 - Characterization}
\end{figure}

Our Cd\textsubscript{3}As\textsubscript{2} nanobelts were grown by chemical vapor deposition (CVD) following a recipe modified from previous reports~\cite{schonherr2015structural, zhang2017controllable}. Briefly, Cd\textsubscript{3}As\textsubscript{2} precursor was heated up to 640$^{\circ}$C and carried downstream by Ar gas flowing between 20-30 sccm to a Si substrate where vapor deposited at 200-250$^{\circ}$C. The growths yielded nanobelts with thickness 80-100 nm, length 10-60 $\mu$m, and width 3-10 $\mu$m as shown in Figure S1(a) in the Supporting Information. Energy dispersive X-ray spectroscopy (EDS) confirmed correct and uniform stoichiometry along the nanobelts [Figure \ref{Figure 1 - Characterization}(b) inset and Figure S1(b)]. Characterization by powder X-ray diffraction (XRD) on an as-grown sample indicated the nanobelts were dominantly the body centered tetragonal centrosymmetric I4\textsubscript{1}/acd (142) phase~\cite{ali2014crystal}, though there might exist a small portion of the nanostructures belonging to the (133) phase implied by a small peak at $2\theta=33\:^{\circ}$~\cite{park2020phase} [Figure \ref{Figure 1 - Characterization}(b)]. 

Field effect transistors (FETs) incorporating single nanobelts were then fabricated [Figure \ref{Figure 1 - Characterization}(a)] by standard e-beam lithography with Cr/Au as contacts. Current vs. source-drain bias (I-V\textsubscript{b}) curves were in general linear. The conductance increased at positive gate voltage (V\textsubscript{g}) [Figure \ref{Figure 1 - Characterization}(d)], indicating an $n$-type channel. The conductance slightly increased at negative V\textsubscript{g} at 78 K, implying the Fermi level can be tuned below the Dirac point.  Field-effect electron mobilities ($\mu$) were up to 3500 cm$^2$/Vs at 78 K [Figure \ref{Figure 1 - Characterization}(c)] in this device (Device \#1) and up to 10$^4$ cm$^2$/Vs in other measured devices (Table S1). The measured $\mu$ is comparable to that in bulk Cd\textsubscript{3}As\textsubscript{2} crystals~\cite{jay1977electron, zdanowicz1983shubnikov} and demonstrates the high quality of the nanobelts. When the temperature is increased from 78 K to 300 K, $\mu$ was about two orders of magnitude lower and the electron concentration was one order of magnitude higher as shown in Figure \ref{Figure 1 - Characterization}(c).

\begin{figure*}
    \centering
    \includegraphics[width=15cm]{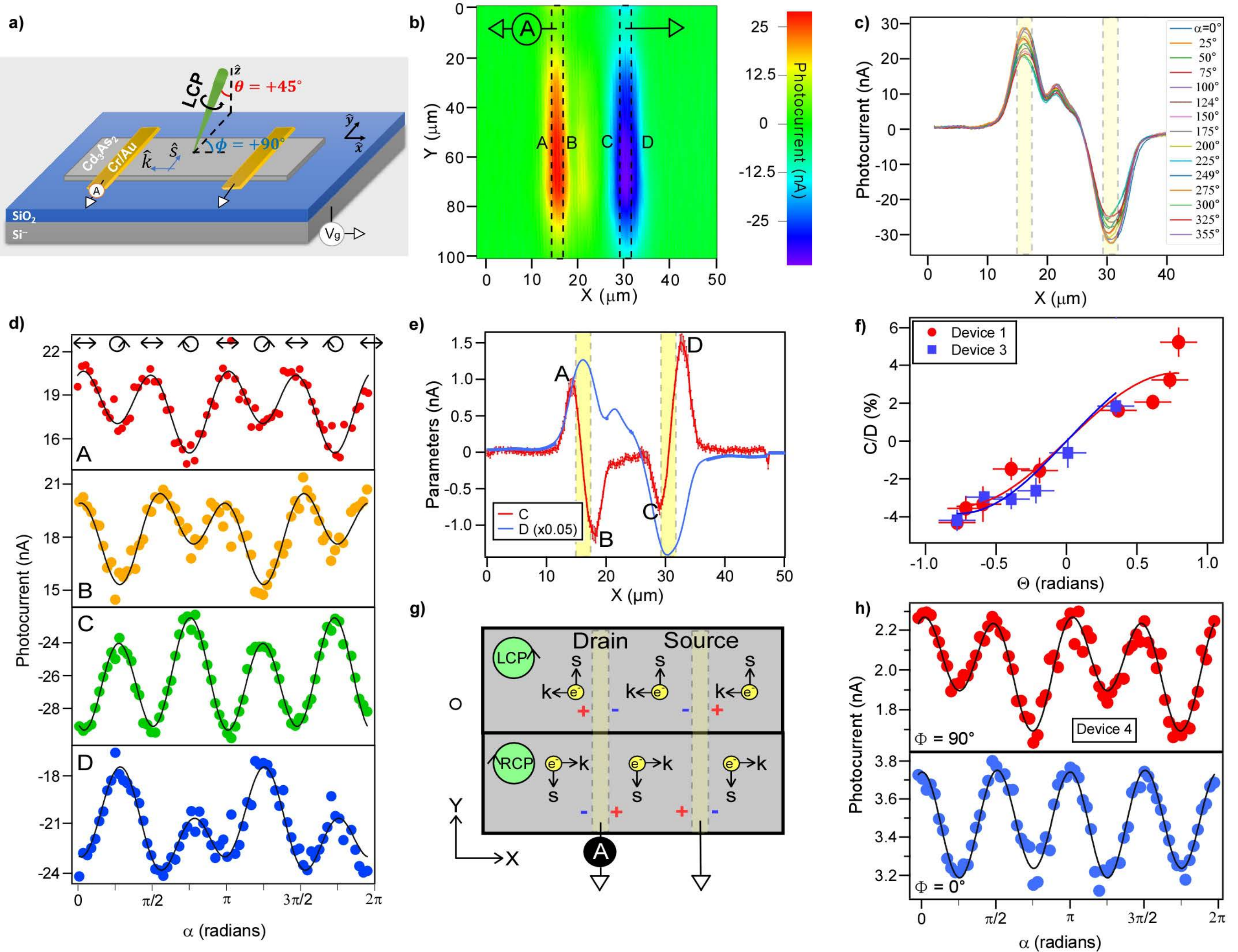}
 
     \caption{Spatially resolved CPGE measurements at room temperature. (a) Schematic of measurement configuration and the coordinate system. The data in (b)-(e) were taken with the laser incident in the y-z plane at $\theta$ = 45$^{\circ}$ and $\phi$ = 90$^{\circ}$ as indicated. (b) Short-circuit photocurrent map. A-D denote positions where the CPGE is strongest. The incident laser power was 455 $\mu$W, corresponding to an intensity of approximately 455 W/cm$^2$. (c) Photocurrent cross sections along the channel at different $\alpha$. Shaded areas indicate the contact positions. (d) Photocurrent at positions A-D as a function of $\alpha$. Solid curves are fittings by Equation \ref{CPGE Eqn}. (e) Cross section of extracted $C$ and $D$ (scaled down) along the channel with positions A-D denoted. (f) $\theta$ dependence of $C/D$ for two representative devices. The solid curve is a fit by Equation S1. The uncertainty in angle of about 7$^{\circ}$ is estimated from the laser cone angle focused by the objective lens. (g) Schematic depicting the polarity of helical photocurrent. Electron spin ($s$) is injected in-plane through oblique incidence. $k$ indicates the momentum locked to the injected spin. + (increase) and - (decrease) indicate the modulation to the photocurrent caused by spin-momentum coupling.  (h) Photocurrent modulation at different $\phi$ values. Note that $\pi$ periodicity vanishes with in-plane spin longitudinal to the channel. (b)-(e) were taken from Device \#2, (f) from \#1 and \#3, (h) from \#4.}
    \label{Figure 2 - SPCM}
\end{figure*}

To study HDPC, spatially resolved optoelectronic measurements were performed at oblique incidence using a 532 nm continuous wave laser focused by a 10$\times$ objective lens, perpendicular to the channel with an incident angle of $\theta$ = 45 $^{\circ}$, as shown in Figure \ref{Figure 2 - SPCM}(a). Photocurrent images were obtained by raster-scanning the laser on the device plane, as shown in Figure \ref{Figure 2 - SPCM}(b). We found that the laser spot on the sample was highly elongated to a 2.5 $\times$ 40 $\mu$m$^2$ area, as determined from the photocurrent image. This elongation is a result of astigmation caused by oblique incidence through the quartz window of the cryostat~\cite{zheng2021astigmatism}. In order to have high spatial resolution along the device channel, we focused the laser with the short axis of the elongated laser spot along the channel. The photocurrent was only large near the contacts and exhibited opposite polarities. The localized photocurrent is different from the nonlocal distribution in Sb doped Bi$_2$Se$_3$ devices~\cite{hou2019millimetre}, presumably caused by the fast carrier recombination in the semimetal. The photocurrent can be generated by photovoltaic (PV) and photo-thermoelectric (TE) mechanisms~\cite{fu2011electrothermal}. The photocurrent sign is consistent with an $n$-type channel where band-bending is upward to the metal contact. The external quantum efficiency (EQE) is quite low, on the order of 0.01\%. However, the internal quantum efficiency (IQE) is about 0.1\%, considering only 10\% of the laser power is absorbed by the material. The low IQE is likely caused by the fast carrier recombination. 

HDPC measurements were performed by passing a linearly polarized beam through a rotating quarter wave plate (QWP). A schematic drawing of the experimental setup is shown in the Supporting Information. The achieved output polarization is a function of the angle ($\alpha$) between the fast axis of the QWP and the incident linear polarization, continuously changing between left circular polarization (LCP) at 45$^{\circ}$ and 225$^{\circ}$, to linear polarization (LP) at 0$^{\circ}$, 90$^{\circ}$, 180$^{\circ}$, and 270$^{\circ}$, and to right circular polarization (RCP) at 135$^{\circ}$ and 315$^{\circ}$ as the QWP is rotated. The photocurrent distributions strongly depended on $\alpha$, as shown in Figure \ref{Figure 2 - SPCM}(c). To obtain spatial resolution, photocurrent maps were taken in 5$^{\circ}$ increments in $\alpha$. Then, the photocurrent as a function of $\alpha$ can be constructed at every pixel in the map as shown in Figure \ref{Figure 2 - SPCM}(d). The photocurrent profiles are fit by the following equation to extract modulation amplitudes.

\begin{equation}
    \centering
    I=C\sin(2\alpha)+L_1\sin(4\alpha)+L_2\cos(4\alpha)+D
    \label{CPGE Eqn}
\end{equation}

\noindent where $C$ represents the amplitude of the HDPC, $L_1$ the linear polarization dependent effects, $L_2$ the reflectance difference at $s$ and $p$ polarizations, and $D$ the polarization independent contributions. The distributions of $C$ and $D$ are shown in Figure \ref{Figure 2 - SPCM}(e), and $L1$ and $L2$ in Figure S5. Here we mainly focus on $C$ which is directly related to the HDPC. 

The magnitude of $C$ becomes small at normal incidence and increases with the laser incident angle, $\theta$ [Figure \ref{Figure 2 - SPCM}(f)]. The cryostat window was removed for the incident angle dependence measurements to avoid the beam astigmation. The $\theta$ dependent helical photocurrent follows reasonably well with the expectation from the polarization dependent optical absorption (the solid curve is the fitting by Equation S1) [Figure \ref{Figure 2 - SPCM}(f)]. Note that here we normalize $C$ by $D$ to account for the laser intensity change caused by realignment at each incident angle, as both $C$ and $D$ are linear with laser intensities up to 500 $\mu$W (Figure S3, S4 in Supporting Information). $C$ vanishes when the laser is parallel to the channel with $\phi$ = 0 $^\circ$ [Figure \ref{Figure 2 - SPCM}(h)]. One of the most striking observations is that $C$ is only large when the laser is focused close to the metal-DSM interface, as shown in Figure \ref{Figure 2 - SPCM}(e). This is different from the previous observations in TIs, where the distributions of $C$ were found to be relatively uniform along the channel~\cite{qu2018anomalous}. Interestingly, $C$ has the same sign at opposite contacts in the device channel, but switches sign outside of the channel. The $C$ distributions and the sign-flipping were highly reproducible in more than 10 devices measured to date.

Now we discuss the possible mechanism for the HDPC in the vicinity of the metal-DSM interface. HDPC can be caused by several mechanisms, including the photo-induced inverse spin Hall effect (PISHE), the circular photon drag effect (CPDE), and the CPGE. We will rule out all but the latter mechanism below. PISHE is induced by the diffusion of spins and requires a focused laser~\cite{tang2021modeling}. In our case, $C$ normalized by intensity does not change much with the laser spot size (Figure S4 in SI), indicating it is unlikely caused by PISHE. Additionally, HDPC disappears at normal incidence [Figure \ref{Figure 2 - SPCM}(f)] where PISHE should be strongest. The CPDE relies on the transfer of both linear and angular momentum from incident photons to electrons and inverts its polarity upon reversal of photon helicity~\cite{glazov2014high}. Unlike the CPGE, the CPDE does not require a broken inversion center to manifest HDPC like the CPGE does, and has been observed in centrosymmetric graphene monolayers~\cite{glazov2014high, zhu2019circular}. Since the HDPC in the Cd\textsubscript{3}As\textsubscript{2} nanobelt is observed only close to the contacts, CPDE is unlikely the dominant mechanism. 

We attribute the observed HDPC to the CPGE as a result of the field-induced symmetry breaking near the metal contact. Because of spin degeneracy of the Weyl nodes in DSMs, the CPGE is expected to vanish. Only when the photoexcitation is close to the contact does the metal interface locally break the inversion symmetry, leading to spin dependent charge transport. This hypothesis is consistent with all experimental results including the $\theta$ and $\phi$ dependence and the spatial distribution of the helical photocurrent. The in-plane spin is flipped at negative $\theta$, leading to momentum reversal and the sign change of the helical photocurrent [Figure \ref{Figure 2 - SPCM}(f)]. $C$ vanishes at $\phi$ = 0 $^{\circ}$ because the injected in-plane spins parallel to the channel do not modulate the measured longitudinal current [Figure \ref{Figure 2 - SPCM}(h)]. The signs of $C$ at different locations are also consistent with the spin-momentum locking picture depicted in Figure \ref{Figure 2 - SPCM}(g). $C$ can be represented by the difference in photocurrent under LCP and RCP photoexcitation, i.e. $\Delta I = I_{LCP}-I_{RCP}$. A LCP photon creates an electron with an in-plane spin corresponding to a momentum along the negative $x$-axis as shown. If generated inside the device channel (position B), this electron tends to move towards the drain connected to the preamp, resulting in a photocurrent smaller than that generated by a RCP photon at this location, leading to a negative $C$. When the laser is moved out of the channel (position A), a LCP photon instead generates an electron with a higher chance of moving away from the drain contact. As a result, the sign of $C$ is flipped. The signs of $C$ for positions C and D are also consistent with this picture. 

\begin{figure}
    \centering
    \includegraphics[width=15cm]{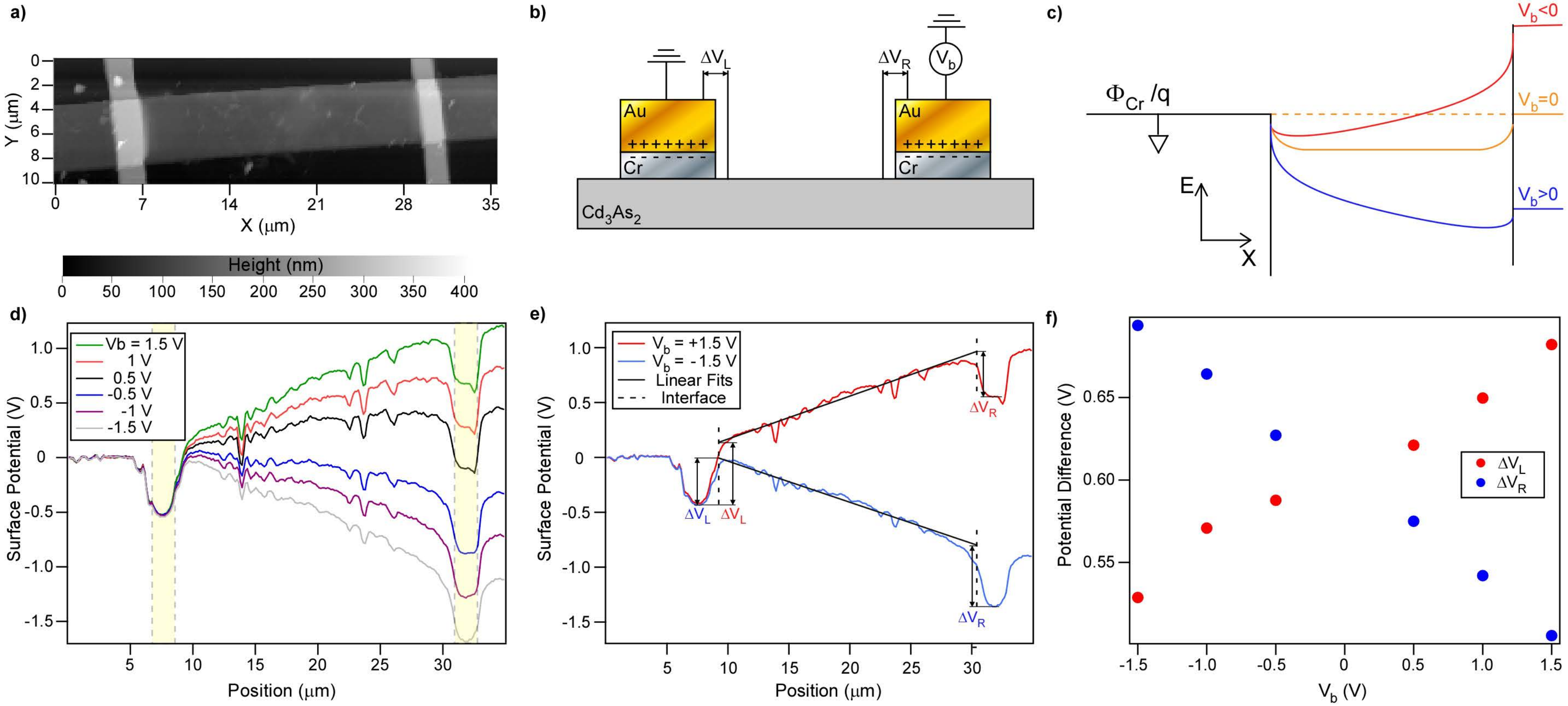}
    \caption{Surface potential measurements by KPFM. (a) Atomic force microscopic (AFM) image of Device \#1. (b) Schematic depicting the device configuration and the definition of $\Delta V$ and the Galvani potential. (c) Band bending diagrams at various $V_b$. (d) Longitudinal cross sections of surface potential taken by KPFM at various $V_b$. (e) $\Delta V$ extracted by extrapolating from a linear fit of the channel profile to the interface. (f) $\Delta V$ as a function of $V_b$ at both left and right contacts.}
    \label{Figure 3 - AFM}
\end{figure}

To confirm the Schottky electric field, we performed Kelvin probe force microscopy (KPFM) under various source-drain bias voltages ($V_b$). The work function of Cr ($\Phi_{Cr}=4.37$ eV) \cite{ossowski2008density} and electron affinity of Cd\textsubscript{3}As\textsubscript{2} ($\chi_{Cd_3As_2}=4.5$ eV) \cite{huang2019high} imply a built-in potential ($V_{bi}$) of approximately $0.13$ eV. An abrupt surface potential change of about 0.6 V can be seen near the contacts [Figure \ref{Figure 3 - AFM}(d)], which is larger than the expected 0.13 eV due to the inclusion of a Galvani potential ($V_{Au-Cr}$) induced at the Au-Cr interface from accumulation of opposite charges~\cite{peljo2016contact} [Figure \ref{Figure 3 - AFM}(b)]. The total surface potential difference measured by KPFM at the contact is hence $\Delta V=V_{bi}+V_{Au-Cr}$. While $V_{Au-Cr}$ is bias-independent due to the large electron density in the metals, $V_{bi}$ depends on the bias. Such bias dependence is readily observed in the KPFM results. $\Delta V$ increases (decreases) at the reversely (forwardly) biased contact, as expected from Schottky junctions [Figure \ref{Figure 3 - AFM}(f)]. The bias dependent $\Delta V$ at both contacts can be quantitatively understood by a back-to-back diode circuit model, using a Schottky barrier height of 0.1 eV (see details in the Supplemental Information). This value is consistent with the expected band bending between Cr and Cd\textsubscript{3}As\textsubscript{2}~\cite{huang2019high}. Using this value and the electron concentration extracted from gate measurements, we estimate a junction electric field of 2-20 MV/m (Table S1 in the Supporting Information), similar to values in previous works on electric field induced CPGE~\cite{dhara2015voltage, ma2019nonlinear}. 

\begin{figure}
    \centering
    \includegraphics[width=12cm]{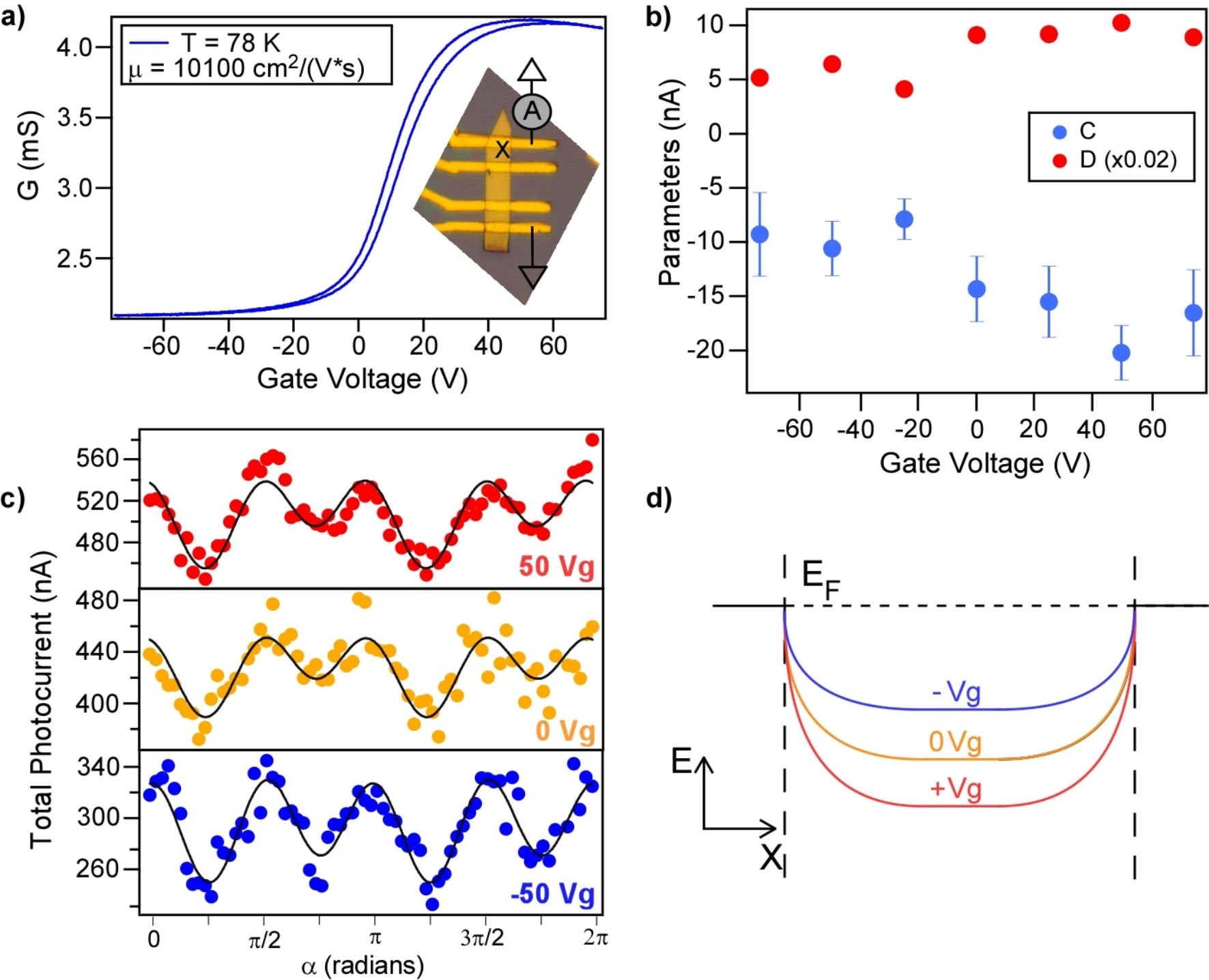}
    \caption{Gate dependent CPGE measured in Device \#2 at 78 K. (a) The conductance as a function of gate voltage. The extracted field-effect mobility is $10^4$ cm$^2$/(V s). Inset: an optical image of the measured device. (b) Extracted $C$ and $D$ at position X as a function of gate voltage. The incident laser power and intensity were similar to that of the 300 K measurement. (c) Photocurrent at position X as a function of $\alpha$ at various gate voltages. The amplitude of the $\pi$ periodic signal noticeably decreases at more negative gate voltages. (d) Schematic depicting the effect of gate voltage on interfacial band bending.}
    \label{Figure 4 - Gate}
\end{figure}

Finally, we performed gate dependence measurements to study the effect of contact band bending on CPGE. To achieve strong gate dependence, the device was cooled down to 78 K. A field effect mobility as high as 10$^4$ cm$^2$/(V s) was extracted in this device (Device \#2). The saturation of gate dependence at both positive and negative $V_g$ as seen in Figure \ref{Figure 4 - Gate}(a) has been previously reported \cite{wang2016aharonov, li2022gate} and was attributed to electronic screening as the Fermi level is tuned away from the Dirac point. Both $C$ and $D$ are about 10 times stronger at 78 K than at room temperature under similar photoexcitation intensity, likely because of higher carrier mobility, longer spin relaxation time, and longer carrier lifetime. Both $C$ and $D$ are strongly modulated by $V_g$ as shown in Figure \ref{Figure 4 - Gate}(b,c). The magnitude of $D$ increases at positive $V_g$, while the magnitude of $C$ first increases at positive $V_g$ but then decreases at $V_g$ = 75 V. The contact band bending can be modulated significantly by $V_g$ as depicted in Figure \ref{Figure 4 - Gate}(d). Larger band bending results in an increase in photocurrent because of more efficient charge separation at the contact. The stronger Schottky field at positive $V_g$ also leads to a larger HDPC, consistent with the field-induced symmetry breaking mechanism. The decrease in HDPC at higher $V_g$ can be understood by the shortened Fermi arc (which hosts the spin-momentum coupled energy states) when Fermi level is shifted away from the Dirac point~\cite{lin2020electric}. 

\begin{figure}
    \centering
    \includegraphics[scale=0.4]{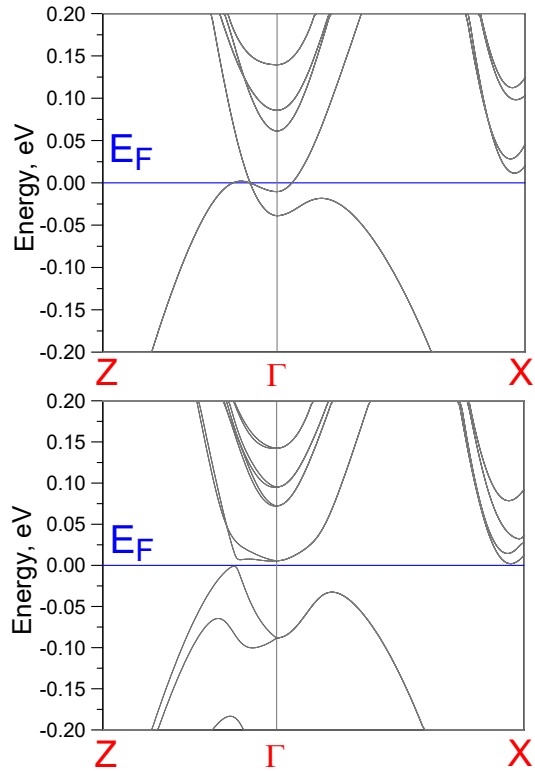}
    \caption{LDA calculated band structure of Cd\textsubscript{3}As\textsubscript{2} without (above) and with (below) a 0.4 \AA \:structural distortion.}
    \label{Figure 5 - LDA}
\end{figure}

To explore the exact mechanism of how the Schottky electric field induces the CPGE, we have conducted a numerical density functional based calculation of the electronic structure of Cd\textsubscript{3}As\textsubscript{2} using local density approximation (LDA) and the full potential linear muffin-tin orbital method~\cite{savrasov1996linear}. The in-plane Schottky electric field is expected to be along the nanobelt growth direction (110)~\cite{park2020phase}, causing some crystallographic distortions along this direction. We have calculated one-electron energy bands by imposing small (0.4 \AA) shifts of Cd cations and As anions in opposite directions. These distortions break the inversion symmetry of Cd\textsubscript{3}As\textsubscript{2}. Therefore, the transition to either the WSM or TI phase is expected. The result of our calculation is shown in Figure \ref{Figure 5 - LDA} where the originally two-fold degenerate bands along $\Gamma$-Z [001] line of the Brillouin Zone are now split due to the imposed distortions, and a small gap is opened at the Fermi level at the original position of the Dirac point. We have analyzed the electronic structure in the vicinity of the gap opening and concluded that the system transitions to the insulating (not WSM) phase.

To better understand this numerical result, we have constructed a $4\times4$ $k\cdot p$ model including an inversion breaking term (details are given in the Supporting Information). Simulations of the $k\cdot p$ model confirm that the inversion broken structure of Cd\textsubscript{3}As\textsubscript{2} along (110) transitions to the TI phase. Although it is not exactly clear whether the electric field in our experiment is sufficiently large to create ion shifts used in our density functional calculation, one can generally expect that in the vicinity of the contact, the nanostructure will acquire some sort of perturbation along the (110) crystallographic line. Therefore, our generic $k\cdot p$ model assumes that the transition from DSM to TI phase should be a robust feature in the vicinity of the contact regardless of the exact cause of inversion breaking.

In summary, we have performed spatially resolved circular polarization dependent photocurrent measurements in Cd\textsubscript{3}As\textsubscript{2} nanobelt FETs. Helicity dependent photocurrent was observed near the metal contacts. We attribute its origin to the Schottky electric field induced inversion symmetry breaking, which is supported by the contact band bending observed by KPFM. At 78 K, gate voltage can significantly modulate the magnitude of the helical photocurrent. A density functional based calculation confirms that Cd\textsubscript{3}As\textsubscript{2} transitions from a DSM to a TI under atomic distortion along the (110) crystallographic line. The study not only provides key insights on better understanding the interfacial effects on topological phases but also offers a novel in-situ control of helicity dependent optoelectronic properties. The experimental approach used in this study to map helicity dependent photocurrent may find applications for identifying topological phases with high spatial resolution.

\section{Experimental Methods}

\quad\: Nanostructures of Cd$_3$As$_2$ were grown via chemical vapor deposition through a vapor-solid growth mechanism~\cite{park2020phase, schonherr2015structural}. Cd$_3$As$_2$ lumps and powder of 99\% purity purchased from Alfa Aesar were placed 10 cm upstream from the center of a quartz tube within a Lindberg tube furnace. The silicon wafer was elevated to the center of the tube using a quartz platform. The wafer was placed 15 cm downstream from the center at a temperature of $200-250^{\circ}$C. The furnace was ramped up to $640^{\circ}$C in 10 minutes and held at that temperature for 35 min before being cooled naturally to room temperature. FET devices incorporating single nanobelts were made on 300 nm SiO\textsubscript{2} covered Si wafers, which served as the back gate. 5 nm Cr / 295 nm Au were deposited by e-beam evaporation as source and drain contacts. As acetone is hygroscopic and Cd\textsubscript{3}As\textsubscript{2} is sensitive to water exposure, only water-free acetone was used to perform lift off.

Optoelectronic measurements were performed at oblique incidence using a 532 nm continuous wave laser focused by a 10$\times$ N.A. 0.25 objective lens. A pair of mirrors mounted on galvanometers raster scanned the laser spot across the entire device. Photocurrents were measured as a function of laser position and photon polarization through a DL1211 preamplifier followed by a LabView data acquisition board. Circular polarization was controlled by rotating a zero-order QWP. A schematic drawing of the setup can be found in Figure S2 in the Supporting Information. Low temperature measurements were performed in a Janis ST-500 optical cryostat. KPFM measurements were performed using Veeco Dimension 3100.

\begin{acknowledgement}

\quad\: We thank Pedro de Oliveira for assistance in KPFM measurements, and Valentin Taufour and Rahim Ullah for assistance in XRD characterization. This work was supported by the U.S. National Science Foundation Grant DMR-2105161 and DMR-1832728. Part of this study was carried out at the UC Davis Center for Nano and Micro Manufacturing (CNM2). Device fabrication was partially carried out at the Molecular Foundry, which is funded by the Office of Science, Office of Basic Energy Sciences, of the U.S. Department of Energy under Contract No. DE-AC02-05CH11231. 

\end{acknowledgement}

\begin{suppinfo}

SEM and EDS characterizations of as-grown samples, a table of device parameters, schematic of experimental setup, more data and analysis of photocurrent dependence on laser intensity, angle of incidence, laser spot size, distributions of L$_1$ and L$_2$, estimation of Schottky barrier height via a circuit model, and simulation of band structures under structural distortion.

\end{suppinfo}

\bibliography{NanoLetters_Style.bbl}

\end{document}